\documentclass[review]{elsarticle}
\usepackage{lineno,hyperref}
\modulolinenumbers[5]
\usepackage{color}
\usepackage{xspace}
\usepackage{graphicx}	
\usepackage{amsmath}	
\usepackage{amssymb}
\makeatletter
\DeclareRobustCommand{\LaTeX}{L\kern-.26em%
        {\sbox\z@ T%
         \vbox to\ht\z@{\hbox{\check@mathfonts
           \fontsize\sf@size\z@
           \math@fontsfalse\selectfont
          A\,}%
         \vss}%
        }%
     \kern-.15em%
    \TeX}
\makeatother
\def\edth{\;\raise1.0pt\hbox{$'$}\hskip-6pt\partial\;}
\def\baredth{\;\overline{\raise1.0pt\hbox{$'$}\hskip-6pt
\partial}\;}
\def\gsim{~\rlap{$>$}{\lower 1.0ex\hbox{$\sim$}}}

\newcommand{\be}{\begin{equation}}
\newcommand{\ba}{\begin{eqnarray}}
\newcommand{\ee}{\end{equation}}
\newcommand{\ea}{\end{eqnarray}}

\newcommand{\abs}[1]{\left\vert#1\right\vert}
\begin{document}
\title{Computing the luminosity distance via optimal homotopy perturbation method}
\tnotetext[mytitlenote]{Fully documented templates are available in the elsarticle package on \href{http://www.ctan.org/tex-archive/macros/latex/contrib/elsarticle}{CTAN}.}

\author[mymainaddress,mysecondaryaddress]{Bo Yu}


\author[mymainaddress]{ Zi-Hua Wang}
\author[liumainaddress,liusecondaryaddress]{De-Zi Liu}
\author[mysecondaryaddress,forzhang]{Tong-Jie Zhang\corref{mycorrespondingauthor}}
\cortext[mycorrespondingauthor]{Corresponding author}
\ead{tjzhang@bnu.edu.cn}

\address[mymainaddress]{College of Mathematical Sciences, Dezhou University, Dezhou 253023,  China}
\address[mysecondaryaddress]{Institute for Astronomical Science, Dezhou University, Dezhou 253023, China}
\address[liumainaddress]{South-Western Institute for Astronomy Research, YunNan University, Kunming650500, China}
\address[liusecondaryaddress]{The Shanghai Key Lab for Astrophysics,Shanghai Normal University ,100 Guilin Road,Shanghai200234,China}
\address[forzhang]{Department of Astronomy, Beijing Normal University, Beijing, 100875, China}
\begin{abstract}
We propose a new algorithm for computing the luminosity distance in the flat universe with a cosmological constant based on Shchigolev’s homotopy perturbation method, where the optimization idea is applied to prevent the arbitrariness of initial value choice in Shchigolev’s homotopy. Compared with the some existing numerical methods, the result of numerical simulation shows that our algorithm is a very promising and powerful technique for computing the luminosity distance, which has obvious advantages in computational accuracy,computing efficiency and robustness for a given $\Omega_m$. 
\end{abstract}

\begin{keyword}
	\texttt{Distance scale\sep Numerical-cosmology\sep Optimal homotopy perturbation method}
\end{keyword}
\maketitle
\section{Introduction}
\label{sec:intro}

Considerable attentions have been paid to the numerical computation of cosmological distances in recent research[1,2,3,4,5,6,7]. One of the most fundamental distances in cosmology is the luminosity distance $d_L$ , which depends on the redshift and the cosmological parmeters. As is well known, the luminosity distance can not be expressed in terms of a simple analytical function of redshift and the parameters of the underlying cosmological models. Taking the general lambda cold dark matter ($\Lambda$CDM), the luminosity distance is only expressible in terms of transcendental functions or elliptic integral functions. Although the computing speed of the state-of-the-art computers is very fast, the distance analysis involving in many studies (e.g. supernova cosmology) still require extensive computions of the numerical integrals, and is time-comsuming[2,6].  Therefore, accurate and efficient strategies of computing  luminosity distance are vitial in modern cosmology.

Up to now, many analytical and numerical approaches have been proposed for avoiding the difficulty of heavy computation pressure. Pen(hereafter Pen99) proposed a simple algebraic fitting formula that has a global relative error of less than 4\%~[1] . Liu et al.~[2]  presented two efficient numerical strategies of calculating elliptic integrals for luminosity distance in flat $\Lambda$CDM models. Wickramasinghe and Ukwatta(hereafter WU10)~[3] obtained a different approximate expression of luminosity distance, which has a smaller relative error than Pen99. Hao et al.~[4]  used the Pad\'e approximant technique to obtain an different analytical expression. Maarten et al. used hypergeometric functions to derive an another analytical formula of luminosity distance for flat $\Lambda$CDM models~[6] .By solving a certain differential equation based on the homotopy perturbation method (HPM), a new way to calculate luminosity distance is proposed by Shchigolev (hereafter Shch17)~[7], which is different from that of previous methods. In general, these evaluation methods of luminosity distance can be classified into two kinds; one is the method based on simplification of elliptic integrals in luminositydistance and the other is the solution of a ceatain  nonlinear differential equation which the luminosity distance satisfies to.

Based on methods of simplification of elliptic integrals simple and efficient formulas over a large range of redshift usually can be obtained. However, methods of simplification of elliptic integrals have no advantage in calculating precision in certain small redshift range. In contrast, a formula for calculating the luminosity distance with less relative error over a small range of redshift can be derived by solving a certain nonlinear differential equation based on the homotopy perturbation method.
The disadvantage of the solution of a ceatain  nonlinear differential equation is that the redshift range of calculating luminosity distance is slightly smaller than that of the method based on simplification of elliptic integrals.

The homotopy perturbation method that takes full advantage of homotopy and perturbation in solving nonlinear differential equations, was first proposed by He [8,9]. It has been shown that a wide range of nonlinear differential equations can be solved to yield an highly accurate solution by homotopy perturbation method with one or two iterations.Later, several modifications of homotopy perturbation method have been quickly proposed, such as optimal homotopy perturbation method[10,11], which can get a reliable approach to nonlinear problems,and optimal homotopy perturbation method coupled with the least squares technology[12,13],and so on.In addition, it is noted that the selection of initial value in Shchigolev's homotopy is arbitrary, meaning that optimization to the selection of initial value can be performed on his method. In this paper, a new algorithm is proposed for the purpose of getting a more accurate and efficient expression of the luminosity distance over a relatively large range of redshift. The rest of the paper is organized as follows. The differential equation which the luminosity distance in flat $\Lambda$CDM models satisfies to is built in Section~2. In  Section~3, the modified algorithm based on Shchigolev’s method is presented. The performance of the proposed algorithm against some existing exact methods are tested in  Section~\ref{sec:met:performance}. Finally, the conclusion is given in  Section~\ref{sec:conclusion}.

\section{Differential equation of luminosity distance for flat Lambda cold dark matter($\Lambda$CDM) universe}
\label{sec:theory}
As mentioned in Liu2011, the luminosity distance $d_L$ in the spatially flat Lambda cold dark matter universe is given by
\begin{equation}\label{eq1}
d_L(z)=\frac{c(1+z)}{H_0} \int_0^z\frac{dt}{\sqrt{\Omega_m(1+t)^3+\Omega_\Lambda}},
\end{equation}
where $\Omega_m$, $\Omega_\Lambda$ are the energy densities corresponding to the matter and cosmological constant, respectively:$\Omega_m +\Omega_\Lambda=1$.

Following the notation in Shch16, we introduce the notation
\begin{equation}\label{eq2}
W(z)=\Omega_m(1+z)^3+\Omega_\Lambda,W(z)|_{z=0}=1.
\end{equation}

Then we rewrite Eq.(\ref{eq1})as:
\begin{equation}\label{eq3}
d_L(z)=\frac{c(1+z)}{H_0} \int_0^z\frac{dt}{\sqrt{W(t)}}.
\end{equation}
\begin{equation}\label{eq4}
\frac{d_L(z)H_0}{c(1+z)}= \int_0^z\frac{dt}{\sqrt{W(t)}}.
\end{equation}

By differentiating the  Eq.(\ref{eq4}), one can get 
\begin{equation}\label{eq5}
\frac{d}{dz}\left[\frac{d_L(z)H_0}{c(1+z)}\right]=\frac{1}{\sqrt{W(z)}}.
\end{equation}
For simplicity sake, we introduce:
\begin{equation}\label{eq6}
1+z=x,u(x)=\frac{d_L(z)H_0}{cx},
\end{equation}
and rewrite equation (\ref{eq5}) as
\begin{equation}\label{eq7}
\left[\frac{du}{dx}\right]{|_{x=1}}=\frac{1}{\sqrt{W(x-1)}}
\end{equation}

Combining the Eq.\eqref{eq2}, \eqref{eq4},\eqref{eq5},\eqref{eq6} and \eqref{eq7},we have
\begin{equation}\label{eq8}
u^{'}(x){|_{x=1}}=1,u(x)|_{x=1}=0.
\end{equation}

According to Eq.\eqref{eq7} and \eqref{eq8},the second derivative of $u(x)$ is equal to
\begin{equation}\label{eq9}
\frac{d^2u(x)}{dx}=-\frac{1}{2}W^{-\frac{3}{2}}(x-1)\frac{dW(x-1)}{dx}.
\end{equation}

Combining the Eq.\eqref{eq8} and \eqref{eq9},one can obtain the Cauchy problem
\begin{equation}\label{eq10}
u^{''}+\frac{1}{2}W^{'}(x-1){u^{'}}^3=0;u^{'}(x){|_{x=1}}=1,u(x)|_{x=1}=0.
\end{equation}
where the superscript $'$ stands for the derivative with respect to $x$, and $W(x-1)|_{x=1}=1$.

\section{Optimal homotopy perturbation method (OHPM)for calculation of luminosity distance}
\label{sec:e&d}
Eq.\eqref{eq10} is a typical nonlinear second-order differential equation, which can be accurately solved by integration method, and its exact result is formula (1). So we will use analytic method to obtain the approximate solution of this equation. In general,the nonlinear differential equation can be solved very well to yield an highly accurate solution by homotopy perturbation method with one or two iterations.Using homotopy perturbation technique, nonlinear differential equations can be transformed into an infinite number of linear (simple)differential equations.

Now since the homotopy perturbation technique has become standard and concise, the reader can refer to[8,9] for its  basic idea. Let us assume that  a series in $p$ can be used to represent the solution of Eq.\eqref{eq10}.The specific expression for the series in$p$ is as follows:
\begin{equation}\label{eq11}
u=u_0 +pu_1+p^2u_2+p^3u_3+\dots,
\end{equation}
where $p \in \left[0,1\right]$ is an embedding parameter. Setting $p=1$ results in the approximate solution of  Eq.\eqref{eq10}.

In according with the procedure of the HPM, we build the homotopy as follows:
\begin{equation}\label{eq12}
u^{''}+c_1+p\left[\frac{1}{2}W^{'}(x-1){u^{'}}^3-c_1\right]=0,
\end{equation}
where $p \in \left[0,1\right]$ ,and the constant $c_1$ is introduced into Eq.\eqref{eq10} with an initial condition $u^{''}+c_1=0$ at $p=0$.

Substituting equation (11) into equation(12), and equating coefficients with the identical powers of $p$, a set of equations are obtained as follows:
\begin{align}
p^0:& {u_0}^{''}+c_1=0,\label{eq13}\\
p^1:& {u_1}^{''}+\frac{1}{2}W^{'}(x-1){{u_0}^{'}}^3-c_1=0,\label{eq14}\\
&\dots\dots\dots\dots\notag
\end{align}

According to Eq.\eqref{eq10}, the initial conditions for $u_i(x)$ can be set as follows:
\begin{align}\label{eq15}
&{u_0}{|_{x=1}}=0, ~{u_0}^{'}{|_{x=1}}=1;\notag\\
&{u_i}{|_{x=1}}=0, ~{u_i}^{'}{|_{x=1}}=1;
\end{align}
whrere $i\geq 1$.

By solving Eq.\eqref{eq13} with the initial conditions(15), one can obtain
\begin{equation}\label{eq16}
u_0=(x-1)-\frac{c_1}{2}(x-1)^2,
\end{equation}

Combining this equation and  Eq.\eqref{eq14}with the initial conditions(15), we get
\begin{equation}\label{eq17}
\begin{split}
u_1=&\frac{1}{2}(x-1)+\frac{c_1}{2}(x-1)^2-\frac{1}{2}\int_1^xW(t-1)\\
&\times{\left[1-c_1(t-1)\right]}^2\{1-c_1\left[4(t-1)-3(x-1)\right]\}dt,
\end{split}
\end{equation}

Substituting Eq.\eqref{eq16} and Eq.\eqref{eq17} into Eq.\eqref{eq11}, the approximate expression for luminosity distance with  unknown constant $c_1$  can be obtained
\begin{equation}\label{eq18}
\begin{split}
\widetilde{d}_L(z)=&\frac{c(1+z)}{H_0}\{0.01c_1^3z^7+(0.028c_1^3-0.042c_1^2)z^6\\
&+(0.021c_1^3-0.126c_1^2+0.063c_1)z^5-(0.105c_1^2-0.21c_1\\
&+0.035)z^4+(0.21c_1-0.14)z^3-0.21z^2+z\},
\end{split}
\end{equation}
where $c$ is the speed of light,$H_0$ is the Hubble constant, $c_1$ is the unknown constant, and $z$ is redshift.

Lastly and most importantly, the unknown constant $c_1$ can be determined by minimizing the raltive error in the approximation of the luminosity distance: 
\begin{equation}\label{eq19}
\bigtriangleup E=\abs{\frac{\widetilde{d}_L-d_L^{num}}{d_L^{num}}}
\end{equation}
where $\widetilde{d}_L$ and $d_L^{num}$ stand for the values of luminosity distance calculated form our approximate expression and the numerical method, respectively.

The implementation of OHPM can be summarized as follows:

Step1: According to the $\Lambda$CDM model, obtain the differential equation Eq.\eqref{eq10} that the luminosity distance $d_L$ should satisfy to;

Step2: Build the homotopy  Eq.\eqref{eq12}; substitute Eq.\eqref{eq11} into Eq.\eqref{eq12}, and then equate coefficients with the identical powers of $p$;

Step3: Solve the set of differential equations consisting of  Eq.\eqref{eq13} and  Eq.\eqref{eq14} that obtained from Step2 with the initial conditions(15), and then get a set of equations with the unknown constant $c_1$;

Step4: According to Eq.\eqref{eq11},Eq.\eqref{eq16} and Eq.\eqref{eq17}, one can obtain the approximate expression for luminosity distance with an unknown constant $c_1$;

Step5: By minimizing  Eq.\eqref{eq19} , the unknown constant $c_1$ can be determined, and then we obtain the approximate expression for luminosity distance.

\section{Performance of OHPM }\label{sec:met:performance}

In this section, the performance of OHPM proposed in Sect.\ref{sec:3} is assessed. The assessment is mainly carried out from two aspects: accuracy and efficiency. In the flat $\Lambda$CDM models, we set $\Omega_m=0.28$ and $\Omega_\Lambda=0.72$,as an example.

\subsection{Accuracy}
\label{sec:4.1}

Figure 1 shows the comparisons of relative error percentages of approximate solutions to $d_L$($\bigtriangleup E$)for different values of $c_1$.
By minimizing  Eq.\eqref{eq19} , the unknown constant $c_1$ is determined to be 0.44274 for the fixed $\Omega_m=0.28$, and then we obtain the approximate expression for luminosity distance.

\begin{figure}[tbp]
	\centering
	\includegraphics[width=8.9cm,height=6.0cm]{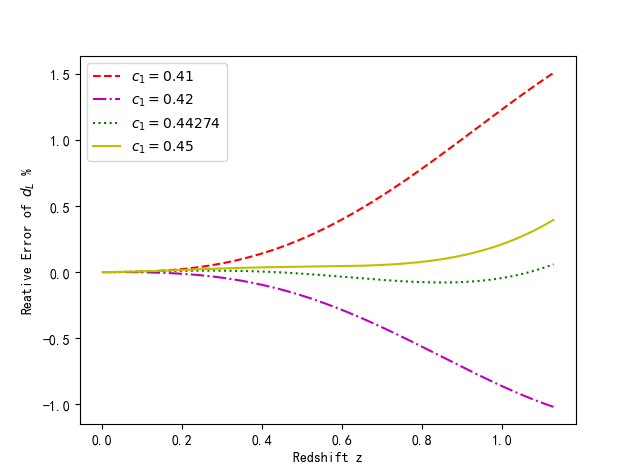}
	\caption{The comparison of approximate solutions to $d_L$ for different constants $c_1$. }
\end{figure}

Using Python, one can obtain a series of numerical solutions to the Eq.\eqref{eq1},and the approximate solutions of Eq.\eqref{eq18}. Compared with some existing methods, the relative error percentages of the numerical and approximate solutions of the same sample are given in Table 1. Seen from Table 1, a best approximation to the exact value of $d_L$ for redshift range $0.1< z\leq 1.129$ can be obtained by our method. For a fixed $\Omega_m=0.28$,it is clear from relative error percentages of $d_L$ in Figure 2  that our method obviously outperforms some existing methods for $0\leq z\leq 1.129$. Relative error percentages of approximate solutions to $d_L$($\bigtriangleup E$)as a function of z for different $\Omega_m$ is shown in Figure 3. From the Figure 4, for any redshift in  $0\leq z\leq 0.5$ the relative error percentages of $d_L$  are between-0.18\% and 0.22\% for $\Omega_m$ within $0.26\leq \Omega_m\leq 0.30$.The global error surface plot is shown in Figure 5.It shows that the error first decreases and then increases when the variation of $\Omega_m$ from 0.26 to 0.3, for $0.26\leq \Omega_m\textless0.28$ the error first decreases and then increases when the variation of $\Omega_m$ from 0.26 to 0.3,and for $0.28\leq \Omega_m\leq 0.30$ the error first increases and then decreases when the variation of $z$ from 0 to 0.5.

\begin{table}
	\centering
	\caption{Relative error percentages of approximate solutions to $d_L$(Errors \%)in cases of OHPM, Shch17,Pen99 and WU10}.\label{tab-RealtiveEorror}
	\begin{tabular}{|l|c|c|c|r|}
		\hline
		z & OHPM &Shch17 &Pen99 & WU10 \\
		\hline
		0.1 & 0.00351 &0.00164 & 0.25934 & 0.25998 \\
		\hline
		0.3 & 0.01067 & 0.04243 & 0.32274 &0.16452\\
		\hline
		0.5 & 0.01192 & 0.17769 & 0.28412 &0.11771\\
		\hline
		0.7 &0.05888 &0.41710 & 0.21090&0.91915 \\
		\hline
		0.9 & 0.07540 & 0.71596 & 0.13634 &0.07626\\
		\hline
		1.1 &0.02791 &0.98781 & 0.07311&0.06598 \\
		\hline
	\end{tabular}
\end{table}

\begin{figure}[tbp]
	\centering
	\includegraphics[width=8.9cm,height=6.0cm]{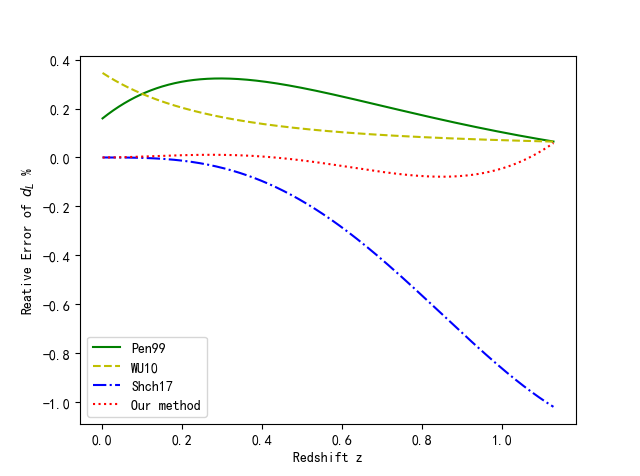}
	\caption{Relative error percentages of approximate solutions to $d_L$($\bigtriangleup E$)as a function of z for $\Omega_\Lambda=0.72$. }
\end{figure}

\begin{figure}[tbp]
	\centering
	\includegraphics[width=8.9cm,height=6.0cm]{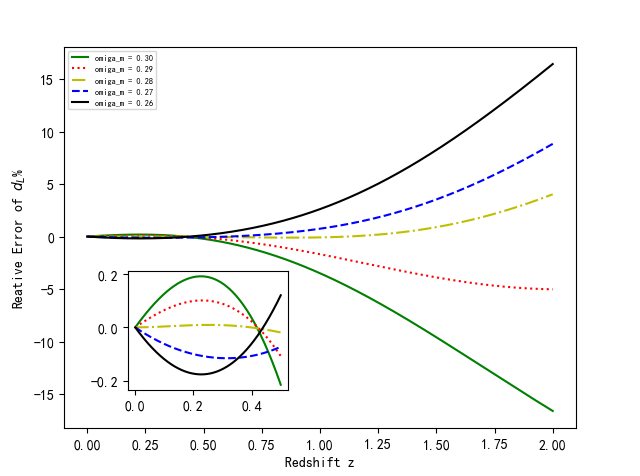}
	\caption{Relative error percentages of approximate solutions to $d_L$($\bigtriangleup E$)as a function of z for different $\Omega_m$.The relative error percentages in $d_L$ for  $0\leq z\leq 0.5$ is amplified,which is shown in the inset.}
\end{figure}

\begin{figure}[tbp]
	\centering
	\includegraphics[width=8.9cm,height=6.0cm]{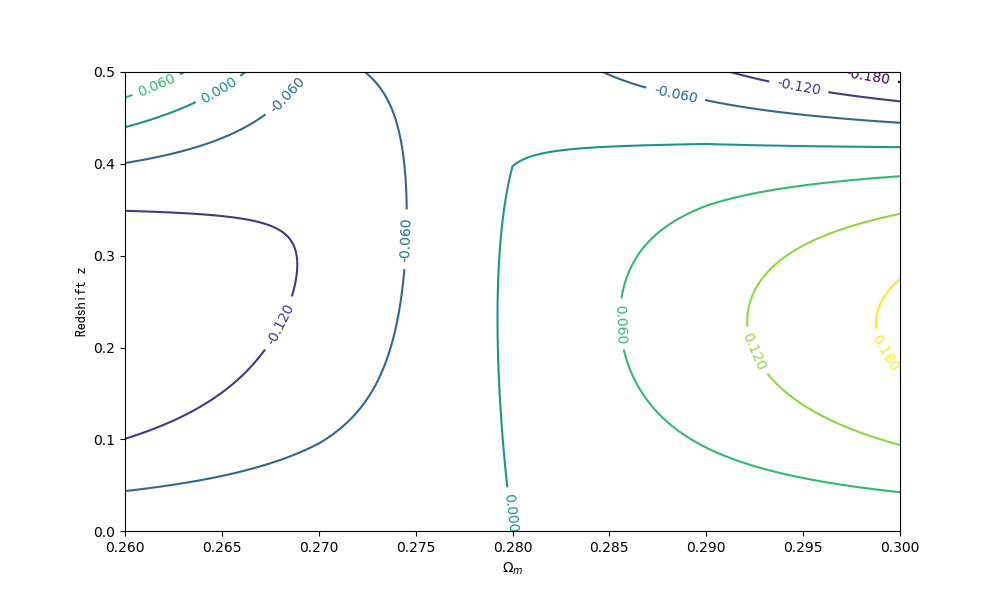}
	\caption{The contour plot for the dietribution of relative error percentages using approximate solutions to $d_L$($\bigtriangleup E$) corresponding to $\Omega_m $ within  $0.26\leq \Omega_m\leq 0.30$. The ``pits'' and ``peaks''in the right region($\Omega_m\geq0.295$) dominate the global error.}
\end{figure}

\begin{figure}[tbp]
	\centering
	\includegraphics[width=8.9cm,height=6.0cm]{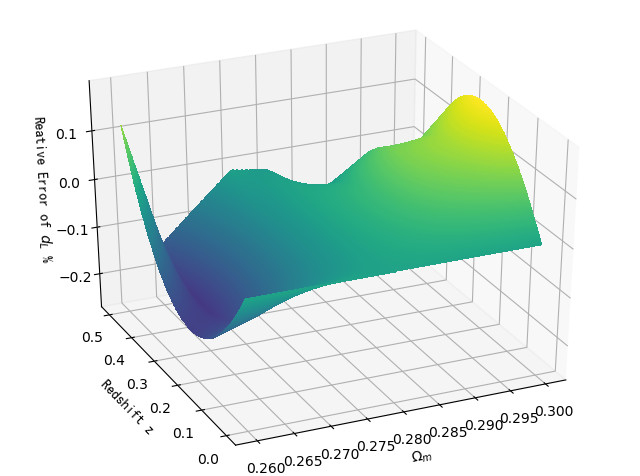}
	\caption{The global error surface plot for approximate solutions to $d_L$($\bigtriangleup E$) corresponding to $\Omega_m $ within  $0.26\leq \Omega_m\leq 0.30$. When the variation of $\Omega_m$ from 0.26 to 0.3, the error first decreases and then increases. When the variation of $z$ from 0 to 0.5, the error first decreases and then increases  for $0.26\leq \Omega_m\textless 0.28$, and the error first increases and then decreases for $0.28\leq \Omega_m\leq 0.30$.}
\end{figure}

\subsection{Efficiency}
\label{sec:4.2}
A comparison of the efficiency of the some exact formulas to calculate the function of $d_L$ is the main purpose of our numerical test. For this purpose, a sample of SN Ia redshifts is created based on the SNAP observation which has 1326 SN data points within 0.1\textless z \textless1.1 (Shafieloo et al)[14].We mock sample has the same redshift distribution as fiducial SNAP, but is 100 times larger in data points than it.Because the calculation accuracy of Shch17 is relatively low, we no longer compare it when comparing the calculation efficiency. So we conducted custom implementations of the methods from Pen99, WU10 and our method in the Python and used its time module. Each implementation of calculating $d_L$ values from the created sample that  contains 132600 SNe redshift points within$ 0.1\leq z\leq 1.1$~is repeated 100 times. 

\begin{figure}[tbp]
	\centering
	\includegraphics[width=8.9cm,height=6.0cm]{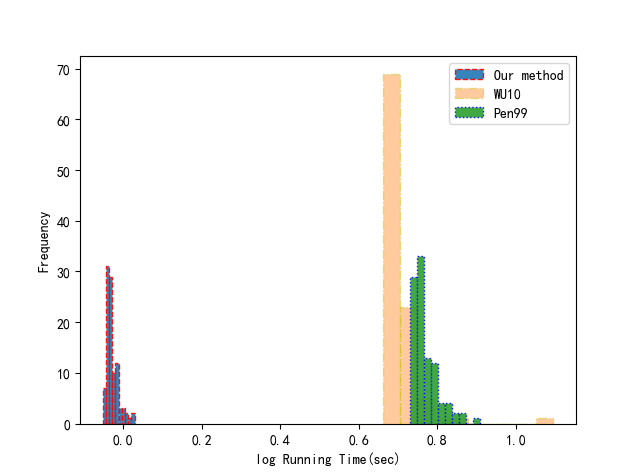}
	\caption{A comparison of the running time of our method,Pen99 and WU10 for redshift range $0.1< z\leq 1.1$. }
\end{figure}

The histogram of the running time in cases of our method and two other methods is shown in Figure 6. Seen from the Figure 6, our method is  obviously faster than that of Pen99 and WU10, and WU10 is slightly faster than Pen99.We note that the results of numerical test may vary depending on the compiler used and the hardware configuration.

\section{Conclusions and discussion}
\label{sec:conclusion}

In this paper, a new algorithm for computing the luminositiy distance for flat universes with a cosmological constan is proposed,which is named OHPM.The proposed algorithm integrates the optimization idea into homotopy perturbation method,where the modified method is applied to prevent the arbitrariness of initial value choice in Shchigolev’s homotopy.

The results of numerical simulation indicate that OHPM has obvious advantages in computational accuracy. The relative error percentages is less than 0.08 percent error for redshit within  $0\leq z\leq 1.129$ for the  fixed $\Omega_m=0.28$. Figure 3,Figure 4 and Figure 5 indicate our algorithm has certain robustness for the different  $\Omega_m$.
In the respects of enhancing computational efficiency, our algorithm possesses great advantage. In addition, OHPM can be extended to other cosmological models.Therefore, OHPM is a very promising and powerful technique to solve the calculation of luminosity distance in theoretical cosmology.

\section*{Acknowledgments}
Bo Yu would like to thank Prof. Jin-Yu He for his kind help.This work was supported by National Key R\&D Program of China (2017YFA0402600) and the National Science Foundation of China (Grants No. 11929301, 11573006).
\bibliographystyle{elsarticle-num}
\bibliography{references}

\begin{thebibliography}{00}
	
	
	
	
	\bibitem[{1}]{paper:01}
	C. Clarkson and C. Zunckel, {Direct Reconstruction of Dark Energy}, Phys. Rev. Lett.
	2010.
	
	
	\bibitem[{2}]{paper:02}
	Pen, Ue-Li, et~al., {Analytical Fit to the Luminosity Distance for Flat Cosmologies with a Cosmological Constant}, Astron.Astrophys.Suppl.Ser. 1999.
	
	\bibitem[{3}]{paper:03}
	M.~Li, X.-D. Li, S.~Wang, Y.~Wang, {Dark Energy}, Communications in Theoretical
	Physics 56~(3) (2011) 525--604.
	
	
	\bibitem[{4}]{paper:04}
	{Liu}, De-Zi and {Ma}, Cong and {Zhang}, Tong-Jie and {Yang}, Zhiliang, {Numerical strategies of computing the luminosity distance}, Mon.Not.R.Astron.Soc. 2011.
	
	\bibitem[{5}]{paper:05}
	{Wickramasinghe}, T. and {Ukwatta}, T.~N., {An analytical approach for the determination of the luminosity distance in a flat universe with dark energy}, Mon. Not. R.Astron.Soc.2010.
	
	
	\bibitem[{6}]{paper:06}
	{Wei}, Hao and {Yan}, Xiao-Peng and {Zhou}, Ya-Nan, {Cosmological applications of Pad{\'e} approximant}, J COSMOL ASTROPART P. 2014.
	
	\bibitem[{7}]{paper:07}
	{Adachi}, M. and {Kasai}, M.,{An Analytical Approximation of the Luminosity Distance in Flat Cosmologies with a Cosmological Constant},Prog. Theor. Phys.2012.
	
	\bibitem[{8}]{paper:08}
	{Baes}, Maarten and {Camps}, Peter and {Van De Putte}, Dries, {Analytical expressions and numerical {Analytical expressions and numerical evaluation of the luminosity distance in a flat cosmology}}, Mon. Not. R. Astron. Soc.2017.
		
		
		\bibitem[{9}]{paper:09}
		{Shchigolev}, V.~K., {Calculating luminosity distance versus redshift in FLRW cosmology via homotopy perturbation method},Gravitation and Cosmology.2017.
		
		
		\bibitem[{10}]{paper:10}
		{He}, J.,{Homotopy perturbation technique},Computer Methods in Applied Mechanics and Engineering.1999.
		
		
		\bibitem[{11}]{paper:11}
		{He}, Ji-Huan, {Some Asymptotic Methods for Strongly Nonlinear Equations}, International Journal of Modern Physics B.2006.
		
		
		\bibitem[{12}]{paper:12}
		{Shafieloo}, Arman and {Alam}, Ujjaini and {Sahni}, Varun and
		{Starobinsky}, Alexei A., {Smoothing supernova data to reconstruct the expansion history of the Universe and its age},Mon. Not. R. Astron. Soc.2006.
			
			
			\bibitem[{13}]{paper:13}
			Nicolae Heris and  Marinca, Vasile, {Optimal Homotopy Perturbation Method for a Non-Conservative Dynamical System of a Rotating Electrical Machine},Ztschrift Für Naturforschung A.2012.
			
			
			\bibitem[{14}]{paper:14}
			Gupta, A. K.  and  Saha Ray, S., {Comparison between homotopy perturbation method and optimal homotopy asymptotic method for the soliton solutions of Boussinesq–Burger equations}, Computers and Fluids.2014.
			
			
			\bibitem[{15}]{paper:15}
			Hayman Thabet and Subhash Kendre, {Modified least squares homotopy perturbation method for solving fractional partial differential equations},Malaya Journal of Matematik.2018.
			
			
			\bibitem[{16}]{paper:16}
			Constantin Bota and Bogdan C{$\vec{a}$}runtu, {Approximate analytical solutions of nonlinear differential equations using the Least Squares Homotopy Perturbation Method}, Journal of Mathematical Analysis and Applications.2016
			
	\end{thebibliography}

	\end{document}